\documentclass[%
 reprint,
 amsmath,amssymb,
 aps,
preprintnumbers
]{revtex4-2}
\usepackage{graphicx}
\usepackage{dcolumn}
\usepackage{bm}
\usepackage{listings}
\usepackage{array,multirow}
\usepackage[font={small}]{caption}
\usepackage{subcaption}
\usepackage[dvipsnames]{xcolor}
\usepackage{colortbl}
\usepackage{adjustbox}
\usepackage{enumitem}  

\begin{document}

\title{Cohesion and excitations of diamond-structure silicon by quantum Monte Carlo: Benchmarks and control of systematic biases}

\author{Abdulgani Annaberdiyev$^{1}$, Guangming Wang$^{1}$, Cody A. Melton$^{2}$,  M. Chandler Bennett$^{3}$, Lubos Mitas$^{1}$}

\affiliation{1) Department of Physics, North Carolina State University, Raleigh, North Carolina 27695-8202, USA}
\affiliation{2) Sandia National Laboratories, Albuquerque, New Mexico 87123, USA}
\affiliation{3) Materials Science and Technology Division, Oak Ridge National Laboratory, Oak Ridge, Tennessee, 37831, USA}


\begin{abstract}
We have carried out quantum Monte Carlo (QMC) calculations of silicon crystal focusing on the accuracy and systematic biases that affect the electronic structure characteristics.
The results show that 64 and 216 atom supercells provide an excellent consistency for extrapolated energies per atom in the thermodynamic limit for ground, excited, and ionized states.
We have calculated the ground state cohesion energy with both {\em systematic and statistical errors} below $\approx$0.05 eV. The ground state exhibits a fixed-node error of only $1.3(2)$\% of the correlation energy, suggesting an unusually high accuracy of the corresponding single-reference trial wave function. 
We obtain a very good agreement between optical and quasiparticle gaps that affirms the marginal impact of excitonic effects.
Our most accurate results for band gaps differ from the experiments by about 0.2 eV. 
This difference is assigned to a combination of residual finite-size and fixed-node errors. 
We have estimated the crystal Fermi level referenced to vacuum that enabled us to calculate the edges of valence and conduction bands in agreement with experiments.
\end{abstract} 


\pacs{Valid PACS appear here}%

\maketitle

\section{Introduction}

Quantum Monte Carlo calculations have been very successful in addressing the challenges of electron-electron correlations in many real materials as well as in important models. 
Real-space QMC samples the particle coordinates and it typically relies on the fixed-node or fixed-phase approximations to avoid fundamental difficulties from negative or complex quantum amplitudes. 
We can perhaps say that the fixed-node/phase QMC has become a ``standard model'' for many-body wave function electronic structure calculations, especially for condensed and periodic systems.
Despite the burden of the fixed-node/phase bias, the QMC methods are evolving into highly accurate approaches that are viable for a number of properties and types of systems. 

QMC calculations of band gaps in periodic systems have been pioneered more than two decades ago~\cite{mitas_quantum_1994, mitas_electronic_1996, fraser_finite-size_1996}. 
At that time, the size of simulation supercells and achieved statistical quality of the results were very limited by the available computational resources. 
Since then, the calculations have 
advanced to strongly correlated systems such as MnO, FeO, NiO,~\cite{wagner_transition_2007, kolorenc_applications_2011, lee_quantum_2004, munoz_comparison_2020} and more complex materials~\cite{wines_first-principles_2020, melton_many-body_2020, du_first-principles_2020, foyevtsova_ab_2014}.
Recent efforts progressed to calculations of defects,  magnetic states, and systems under strain~\cite{santana_quantum_2017, rodrigues_identifying_2020, azadi_equation_2020, driver_quantum_2010}. 

Recently, larger supercells have been employed in more extensive calculations of promotion and quasiparticle gaps, cohesion energies, and other quantities for previously studied semiconductor systems~\cite{yang_electronic_2020, hunt_quantum_2018}.
Most of the ground state (GS) calculations reaffirmed the accuracy of the QMC results, however, consistent high-accuracy for excited states (EX) proved to be rather laborious. 
In particular, some band gaps appeared to be overestimated with a possible culprit being finite-size effects. 
However, other reasons could not be ruled out either, such as slow and nonmonotonous convergence of total energy and/or different rates of convergence for kinetic vs potential energy components.
Further considerations that complicate accurate estimations involve basis set effects, methods to generate single-particle orbitals, and ultimately, fixed-node (FN) errors. 

Here we present extensive calculations and insights into several of these issues.
For the sake of comparison with previous results, we study the Si solid in diamond structure. Another reason for this choice is the fact that the fixed-node errors in Si systems with single bonds 
appear to be appreciably small~\cite{rasch_communication_2014, wang_binding_2020}. 
The intention is to provide more transparency to enable one to clearly understand the rest of the systematic biases. 
We pay significant attention to certain aspects of finite-size errors that complicate QMC studies in a major way.
It is fair to say that these errors are not fully understood despite a number of thorough previous studies~\cite{williamson_elimination_1997, kent_finite-size_1999, chiesa_finite-size_2006, drummond_finite-size_2008, holzmann_theory_2016}.
Note that this is not only the case for many-body methods. In fact, even in density functional theory (DFT) and post-DFT approaches, this is still a subject of substantial effort, despite decades of dedicated research, see Refs.~\cite{perdew_understanding_2017, lany_assessment_2008, mcclain_gaussian-based_2017, alfe_diamond_2004, gruneis_second-order_2010} and references therein. 
We probe for the agreement between the band gap calculations through promotion
(optical) 
vs quasiparticle gaps using the differences between the cation (CA), anion (AN), and neutral systems.
Instead of introducing new or more sophisticated corrections, we focus on some rather ordinary aspects of such calculations and how they can affect the results. 
Lastly, the obtained gaps combined with an estimation of the Fermi level (FL) are employed to derive the ionization potential and electron affinity of the Si solid. Overall, and 
not very surprisingly, we find that desirable increases in accuracy and statistical resolution of the results require correspondingly
thorough effort to better understand and analyze the systematic errors inherent to QMC methods. 

The paper is structured as follows.
Section~\ref{section:methods} describes the general methodology and possible sources of systematic biases related to QMC calculations.
In Sec.~\ref{section:results_data}, the results, data, and analysis are presented.
Section~\ref{section:conclusions} includes the conclusions and discussions.

\section{Methods}
\label{section:methods}

\subsection{QMC methods and trial functions}
For calculations we use variational Monte Carlo (VMC) and fixed-node diffusion Monte Carlo (DMC) methods in their commonly used formulations~\cite{foulkes_quantum_2001, kolorenc_applications_2011}. 
We employ single-reference Slater-Jastrow trial wave functions 
with Jastrow factors that include one-body \mbox{($eI$)}, two-body \mbox{($ee$)}, and three-body \mbox{($eeI$)} terms.
One exception appears in probing the effect of proper symmetry for the open-shell singlet excited state where we tested two configurations, as explained later. 
The orbitals were calculated by Hartree-Fock (HF) and DFT methods that included hybrid functionals.
The calculations are labeled as QMC/DFT where the first acronym denotes the corresponding QMC approach while the second acronym refers to the method used to generate the corresponding orbitals.

The Jastrow factors were optimized for ground states, and we verified that reoptimization in excited states with single-electron promotion produced negligible changes.
We used $\textsc{qwalk}$~\cite{wagner_qwalk_2009} and $\textsc{qmcpack}$~\cite{kim_qmcpack_2018, kent_qmcpack_2020} for the various QMC calculations
and \textsc{nexus}~\cite{krogel_nexus_2016} for workflow management.
The T-moves algorithm as implemented in  $\textsc{qwalk}$~\cite{casula_beyond_2006} and $\textsc{qmcpack}$~\cite{casula_size-consistent_2010}
was used for DMC calculations so that the resulting energies were variational.
The Si crystal was represented by a periodic supercell with a potential energy given by the well-known Ewald expression~\cite{ewald_berechnung_1921, leeuw_electrostatic_1979, foulkes_quantum_2001}.
The charged supercells have been calculated with a neutralizing background to ensure the convergence of the corresponding Ewald sums.

We have chosen a very conservative time step of 
\mbox{0.0025 Ha$^{-1}$} to avoid extrapolations (note that even larger time steps were shown to have only negligible impact on Si solid energies~\cite{li_cohesive_1991}). We used repeated independent runs to probe for walker population bias on the resulting energies and error bars as it is pertinent for DMC of larger systems~\cite{nemec_diffusion_2010}.
The detailed information about the time-step and walker population biases can be found in Supplemental Material~\cite{supplemental}.

Throughout the paper, we show one standard deviation as the statistical error (in parenthesis). 
In some cases, the errors are given with two digits in order to keep the same number of significant digits for all presented energies.

\subsection{Sources of systematic biases}
One of our goals was to shed more light on the systematic errors involved in QMC calculations. This aspect is becoming more prominent as the accuracy of QMC calculations increases. It calls 
for a more thorough look at sources
of possible biases that could compromise 
the quality of QMC outcomes. In what follows,
we identify the origins of possible biases and outline
some of the choices we have made in order to address these. Further analysis is presented in the results section. 


{\em Accuracy of valence-only Hamiltonians.}
The efficiency of QMC calculations is significantly improved when the cores of heavier atoms are replaced by effective core potentials (ECPs). Of course, that requires verification and testing of fidelity of valence  vs.  all-electron Hamiltonians. 
Here we use the recently generated Si atom correlation consistent ECP (ccECP)
that has been tested on molecular systems such as hydride, oxide, and dimer molecules~\cite{bennett_new_2018}.
Corresponding exact atomic valence energies have been analyzed previously in detail
\cite{annaberdiyev_accurate_2020} as well.
Additionally, we have carried out comprehensive accuracy tests for molecular silicon systems (Si$_x$H$_y$)~\cite{wang_binding_2020} that have similar bonding patterns as the Si crystal structure. The sizes of these systems enabled us to
benchmark the fixed-node biases of single-reference trial functions  using several high-level correlated wave function approaches such as coupled cluster (CC) and configuration interaction (CI) methods as well as CI using a perturbative selection made iteratively (CIPSI)~\cite{huron_iterative_1973} with PT2 corrections. Combined with extensive basis sets 
we were able to obtain total energies and differences (excitations in singlet and triplet channels) within about \mbox{0.025 eV} residual uncertainties. We point out that  this is better than chemical accuracy roughly by a factor of two. We also verified that our results were 
 on par with the best available all-electron state-of-the-art studies~\cite{feller_theoretical_1999, haunschild_new_2012}.
 The remaining bias of about 25 meV corresponds to the discrepancies seen in Si$_2$ and SiO molecules that were studied in generating the ccECP table
\cite{bennett_new_2018}.
We conclude that the ccECP for this regime of binding and excitations is highly accurate and represents the valence energy differences with very high accuracy (see Ref.~\cite{wang_binding_2020} for further discussion). 
This opens a path to address the rest of the errors specified above.

{\em Accuracy of basis sets.} 
We address this aspect by exploring two complementary routes.
One is based on \textsc{crystal} code~\cite{dovesi_quantum-mechanical_2018}  using 
triple-zeta valence with polarization (TZVP: $[3s,3p,1d]$)
gaussian basis set with accurate contractions (see Supplemental Material~\cite{supplemental} for the actual data).
The other option is based on \mbox{\textsc{quantum espresso}} ({QE}) code~\cite{giannozzi_quantum_2009} with plane waves and \mbox{100 Ry} kinetic energy cut-off (more details on convergence  in plane wave energy cut-off can be found in the Supplemental Material~\cite{supplemental}). 

{\em Type of single-particle orbitals and corresponding effective one-particle Hamiltonian.} 
Here we have probed the cases of GGA with PBE functional \cite{perdew_generalized_1996}, Hartree-Fock orbitals, as well as PBE0 \cite{perdew_rationale_1996} orbitals. The detailed analysis is presented in Sec.~\ref{section:results_data} (Results).

{\em Form and optimization of Jastrow factors and their consistency.} 
It is well known that the optimization bias can affect the projection of the nonlocal terms and thus induce presumably small shifts in energy, density, etc. 
Although we expect these effects to be small, one has to verify that this is indeed the case.

{\em Periodicity and finite size biases.} 
The basic finite size model we use for the ground and excited states with promoted single electron is given as follows:
\begin{equation}
    \label{eqn:gs_scaling}
    E_{N}^{GS}=e_0 n+A^{GS}+g^{GS}(N)
\end{equation}

\begin{equation}
    \label{eqn:ex_scaling}
    E_{N}^{EX}=e_0 n+E_g+A^{EX}+g^{EX}(N)
\end{equation}
where $e_0$  is the asymptotic energy per atom, $N$ is the number of electrons, $n$ is the number of chemical formula units (number of atoms in this case),  $A^{GS}$ and $A^{EX}$ are energy offsets, and $E_g$ is the excitation energy. 
Functions  $g^{GS}(N)$, $g^{EX}(N)$ capture finite size effects of the lower order that vanish in the limit $N \to \infty$ (as $n \to \infty$). 
In general, the commonly used form is a reciprocal power term $B/n^{\alpha}$ where $B$ is some constant~\cite{melton_many-body_2020}. 
Unfortunately, this form is rather crude and does not capture the nonlinear effects that come into the consideration such as different behavior of kinetic and potential energies on $n$, the impact of Ewald sums, effects from compensating
background in ionized cells, etc.
Therefore, we also probe energies per atom/chemical formula since for intensive quantities the contaminating terms vanish correspondingly faster. For intensive quantities we used linear extrapolations considering only the two largest calculated sizes since we found
that more general nonlinear fits were not very useful.
The inclusion of smaller supercells provided  very little benefit and indeed made the analysis more complicated with much lower robustness and transparency overall.
Our only simplifying assumption is that depending on the system, dimension and electronic state, the exponent $\alpha>0$ of the subleading term is qualitatively assumed to be at least 1 or close 
to it.  
The generalization of these expressions for charged supercells is
straightforward and it is discussed further in the Results section. 


{\em Fixed-node bias.}
We discuss this aspect throughout the paper. We have shown before~\cite{rasch_communication_2014,annaberdiyev_accurate_2020,bennett_new_2018} that Si systems with closed-shells and single bond patterns exhibit some of the lowest fixed-node errors observed in QMC calculations, typically within 1-2 \% of the correlation energy. The same is true also for the Si atom which shows a bias of only $\approx$ 1.5 \%. This provides a favorable setting for insights into the other systematic errors involved. 

\section{Results}
\label{section:results_data}

\subsection{HF energies and basis sets}

In Table.~\ref{tab:scf_qmc_HF_wfn}, we show VMC energies per atom for HF orbitals calculated by  \textsc{crystal} with Gaussians vs \textsc{{QE}} with orbitals expanded in plane waves for sizes that span the primitive cell, 8, 64, and 216 atom cubic supercells. We report the ground state with 
$\Gamma-$point occupation (GS) as well as the $\Gamma \to \Gamma$
excited state formed by single particle promotion ($\Gamma\Gamma$). 
Clearly, the plane-wave basis set is more accurate showing uniformly lower energies.
Further experimentation with basis sets using \textsc{crystal} has produced only marginal gains that were not able to match the
accuracy of plane waves in \textsc{{QE}}.
{
We note that in principle, one should be able to reach the same complete basis set (CBS) limit using localized gaussian basis sets and adequate computational tools.
However, in practice, the convergence with large basis sets can be challenging so that achieving the CBS limit might be either limited by the used software or impractical. (The difficulties are typically rooted in near-linear dependencies from small exponent gaussians that complicate the stability of the diagonalization.)
}
Therefore, for the rest of the calculations we use the plane wave basis.

\begin{figure}[!htbp]
\centering
\caption{
Discrepancy between \textsc{{QE}} self-consistent HF and uncorrelated VMC/HF energies per atom. The plot 
shows linear extrapolations to the thermodynamic limit in $1/n$
 using $8-64$ atoms (dashed line), and $64-216$ atoms (solid line).
See text for further details.
}
\label{fig:scf_qmc}
\includegraphics[width=1.00\columnwidth]{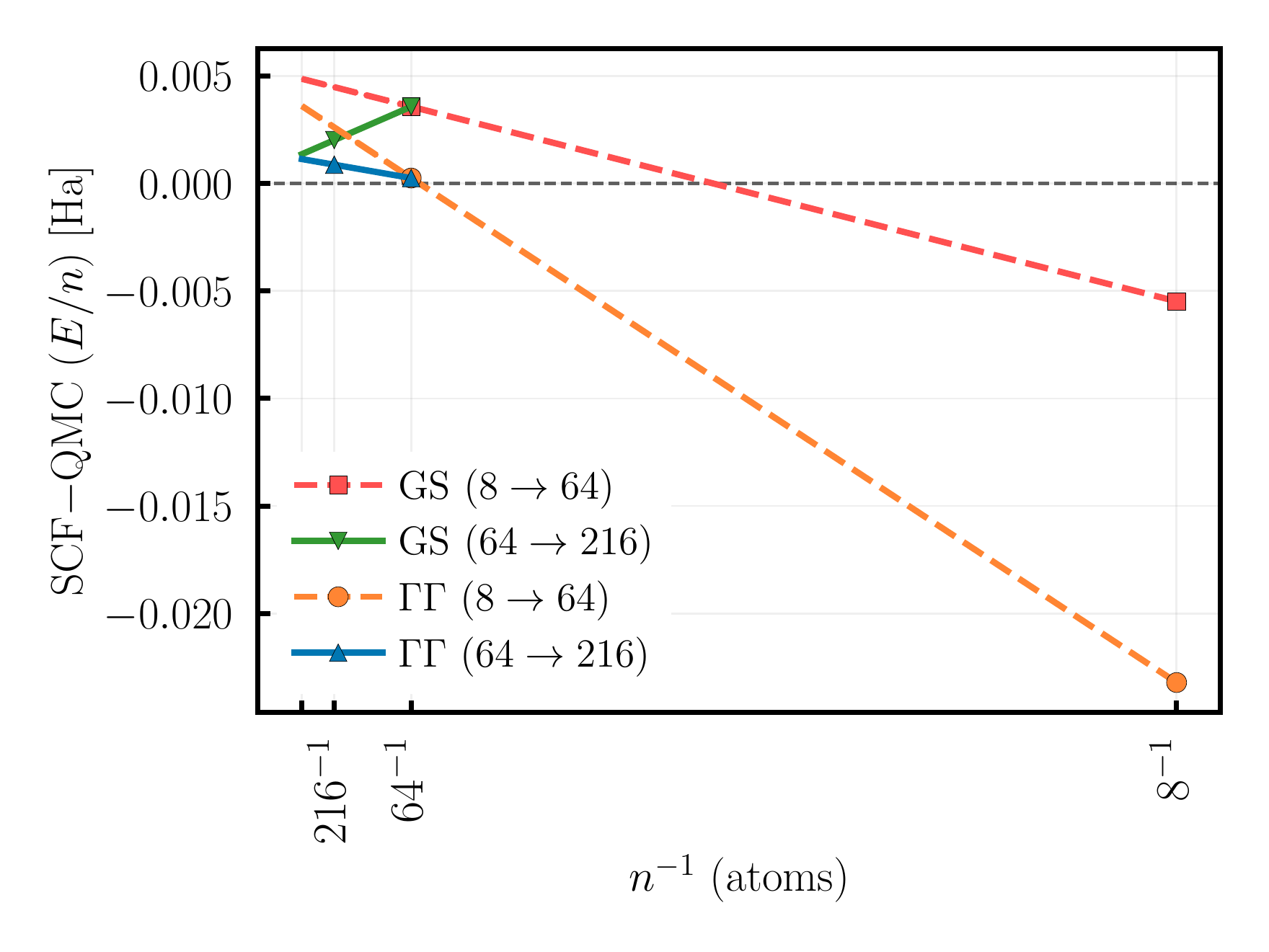}
\end{figure}

We also point out the consistency of the results using linear extrapolations as illustrated in Fig.~\ref{fig:scf_qmc}.
The figure shows two linear extrapolations with dashed lines corresponding to $8-64$ atom supercells while the solid lines correspond to $64-216$ atom supercells.
Note the significant biases for $8-64$ estimators with regard to the reference value.  
In addition, the slope of ground state 
extrapolations changes the sign
while this is not observed for 
the excited state making thus 
any nonlinear extrapolation
very questionable.
A very clear improvement is obtained in extrapolations using $64-216$ supercells with the residual difference being about \mbox{1.4 mHa/atom} from the reference HF energy. A minor difference is not unexpected considering the difference between the methods and some remaining impact from finite sizes.

\begin{table}[!htbp]
\centering
\caption{
VMC energies [Ha] per atom for the supercell $k=\Gamma$ point for GS and $\Gamma$ to $\Gamma$ excitation ($\Gamma\Gamma$) using
the single-reference HF trial function (no Jastrow functions). 
$\infty_{64 \rightarrow 216}$ represents the extrapolated energy using the 64 and 216 atom supercells with $1/n$ extrapolation.
Energies per atom using the corresponding self-consistent field (SCF) codes at high $k-$meshes are also shown.
}
\label{tab:scf_qmc_HF_wfn}
\begin{tabular}{c|ll|ll}
\hline
\rowcolor{lightgray!30}[2pt]& \multicolumn{2}{c|}{\textsc{crystal} HF orbitals} &  \multicolumn{2}{c}{\textsc{{QE}} HF orbitals}\\
\rowcolor{lightgray!30}[2pt]Atoms ($n$) & GS & ${\Gamma\Gamma}$ & GS & ${\Gamma\Gamma}$\\      
\hline
2       &   -3.5718(1) &   -3.5339(2) &    -3.5904(1) &   -3.5416(1) \\
8       &  -3.77535(3) &  -3.75884(3) &   -3.78330(7) &  -3.76558(6) \\
64      &  -3.78529(4) &  -3.78217(5) &   -3.79235(4) &  -3.78903(3) \\
216     &  -3.78381(2) &  -3.78266(3) &  -3.790797(9) &  -3.78965(1) \\
& & & & \\
$\infty_{64 \rightarrow 216}$ 
        &  -3.78319(4) &  -3.78287(5) &   -3.79014(2) &  -3.78992(2) \\
SCF     &  -3.78240    &              &   -3.78878    &              \\

\hline
\end{tabular}
\end{table}

\subsection{Total energies and orbitals}

\begin{table*}[!htbp]
\centering
\caption{
QMC total energies [Ha] at $k=\Gamma$ for $n=[8,64,216]$ atoms with HF, PBE0, and PBE orbitals.
Calculations for Slater trial wave function (WFN) without Jastrow (S) and Slater-Jastrow WFN (SJ) are shown.
}
\label{tab:totals}
\begin{tabular}{ccll|lll}
\hline
\rowcolor{lightgray!30}[2pt] State & Method & WFN & Orbitals & $n=8$ & $n=64$ & $n=216$ \\
\hline

GS             & VMC & S  & PBE  &   -30.2354(6) &  -242.2753(27) &  -817.0897(23) \\
GS             & VMC & S  & PBE0 &   -30.2478(5) &  -242.4541(18) &  -817.8381(14) \\
GS             & VMC & S  & HF   &   -30.2664(6) &  -242.7106(26) &  -818.8121(20) \\
\hline
GS             & VMC & SJ & PBE  &   -31.1587(5) &  -251.1578(13) &  -848.1912(12) \\
GS             & VMC & SJ & PBE0 &   -31.1609(8) &   -251.1697(4) &   -848.2195(5) \\
GS             & VMC & SJ & HF   &   -31.1479(4) &   -250.9287(7) &   -847.2472(8) \\
\hline
GS             & DMC & SJ & PBE  &   -31.2457(9) &  -251.5702(22) &  -849.2468(35) \\
GS             & DMC & SJ & PBE0 &   -31.2440(9) &  -251.5744(13) &  -849.2688(30) \\
GS             & DMC & SJ & HF   &   -31.2370(9) &  -251.4369(18) &  -848.7195(44) \\
\hline
$\Gamma\Gamma$ & VMC & S  & PBE  &   -30.1028(5) &  -242.0868(25) &  -816.8788(27) \\
$\Gamma\Gamma$ & VMC & S  & PBE0 &   -30.1155(4) &  -242.2576(10) &  -817.6114(13) \\
$\Gamma\Gamma$ & VMC & S  & HF   &   -30.1246(5) &  -242.4981(19) &  -818.5652(22) \\
\hline
$\Gamma\Gamma$ & VMC & SJ & PBE  &   -31.0439(4) &  -251.0227(10) &  -848.0547(16) \\
$\Gamma\Gamma$ & VMC & SJ & PBE0 &   -31.0458(2) &   -251.0358(3) &   -848.0805(7) \\
$\Gamma\Gamma$ & VMC & SJ & HF   &   -31.0284(2) &   -250.7860(7) &   -847.0966(7) \\
\hline
$\Gamma\Gamma$ & DMC & SJ & PBE  &  -31.1344(10) &  -251.4338(25) &  -849.1081(31) \\
$\Gamma\Gamma$ & DMC & SJ & PBE0 &   -31.1360(4) &  -251.4371(10) &  -849.1225(25) \\
$\Gamma\Gamma$ & DMC & SJ & HF   &  -31.1233(10) &  -251.2869(16) &  -848.5496(30) \\

\hline
\end{tabular}
\end{table*}

\begin{table*}[!htbp]
\centering
\caption{
QMC/PBE total energies [Ha] at supercell $k=\Gamma$ for charged cases presented as raw data. 
Cation (CA) is obtained by one electron removed from the $k=\Gamma$ state.
AN($\Gamma$/X) represents an extra electron added to the conduction band at $k=\Gamma$/X point. 
}
\label{tab:qmc_charged}
\begin{tabular}{lcl|lll}
\hline
\rowcolor{lightgray!30}[2pt] State & Method & WFN & $n=8$ & $n=64$ & $n=216$ \\
\hline

CA             & VMC & S  &  -30.4965(4) &  -242.5156(27) &  -817.3213(19) \\
CA             & VMC & SJ &  -31.3905(3) &   -251.3755(4) &   -848.4080(9) \\
CA             & DMC & SJ &  -31.4771(9) &  -251.7891(22) &  -849.4721(30) \\
\hline
AN($\Gamma$)   & VMC & S  &  -29.8442(3) &  -241.8537(24) &  -816.6532(17) \\
AN($\Gamma$)   & VMC & SJ &  -30.8153(3) &   -250.8053(5) &   -847.8367(8) \\
AN($\Gamma$)   & DMC & SJ &  -30.9056(9) &  -251.2213(24) &  -848.9063(37) \\
\hline
AN(X)          & VMC & S  &  -29.9305(3) &  -241.9467(23) &  -816.7409(20) \\
AN(X)          & VMC & SJ &  -30.8992(3) &   -250.8896(8) &   -847.9204(8) \\
AN(X)          & DMC & SJ &  -30.9917(9) &  -251.3005(23) &  -848.9833(38) \\

\hline
\end{tabular}
\end{table*}

Table.~\ref{tab:totals} shows the QMC total energies for HF, PBE0, and PBE orbital sets.
Complete results are listed with VMC for both uncorrelated Slater (S) only and Jastrow-correlated (SJ) trial wave functions as well as fixed-node DMC values. 
The question of orbitals is crucial for QMC calculations since they determine the fixed-node errors and often do have significant impact on the results. 
We can see that for uncorrelated Slater wave functions (S), HF obtains the lowest energies for all states and sizes by large margins.
However, when the correlation is included, the DFT orbitals result in the lowest energies in both VMC and DMC methods.
Similar behavior has been observed a number of times previously, for instance, see Fig. 3 in Ref.~\cite{melton_many-body_2020}. 
Although the lowest total energies are obtained using the PBE0 orbitals, the PBE orbitals lead to very similar total energies indicating a comparable quality of the trial wave function. 
Related results were reported before, showing that even plain LDA orbitals are very close to VMC natural orbitals and lead to similar 
resulting energies~\cite{kent_quantum_1998}.

It is revealing to consider the total energy gains for the DFT orbitals vs the HF orbitals for different sizes. For the largest supercell, DMC/HF results in significantly {\em higher} total energies ($\approx0.5$ Ha). This is very significant especially in relation to the corresponding VMC/HF (S) energy that is {\em lower} by almost 1 Ha. 
{
Additionally, we observed that the gaps from QMC/HF are notably higher than QMC/DFT. 
}
Clearly, the inclusion of correlation reveals  that there are significant differences between these two orbital sets. 
Interestingly, this effect grows with size and it becomes obvious only for larger supercells, while for the smallest one with eight atoms the effect is not clearly discernible.
For instance in GS, the energies per atom for \mbox{DMC/PBE} and \mbox{DMC/HF} differ by \mbox{0.030 eV} for $n=8$, and \mbox{0.066 eV} for $n=216$.

For the sake of completeness, we list here also
the results for charged supercells that are further elaborated on later.
Table.~\ref{tab:qmc_charged} provides QMC/PBE total energies for charged cases.
The cation state has one electron removed  from the highest occupied $\Gamma$ state. We calculate two anion states,
with an added electron to the lowest unoccupied orbitals in $\Gamma$ and X $k$ points. These results are further analyzed in detail in Sec. E.

\begin{figure}[!htbp]
\centering
\caption{
DMC/PBE total energy per atom extrapolations using $64-216$ atoms at the supercell $k=\Gamma$ occupation.
AN($\Gamma$)/AN(X) represents the anionic state with an electron added to the state that corresponds to the primitive cell $\Gamma$/X $k-$point.
Note the excellent agreement of $e_0$ at TDL.
}
\label{fig:pbe_tdl}
\includegraphics[width=1.00\columnwidth]{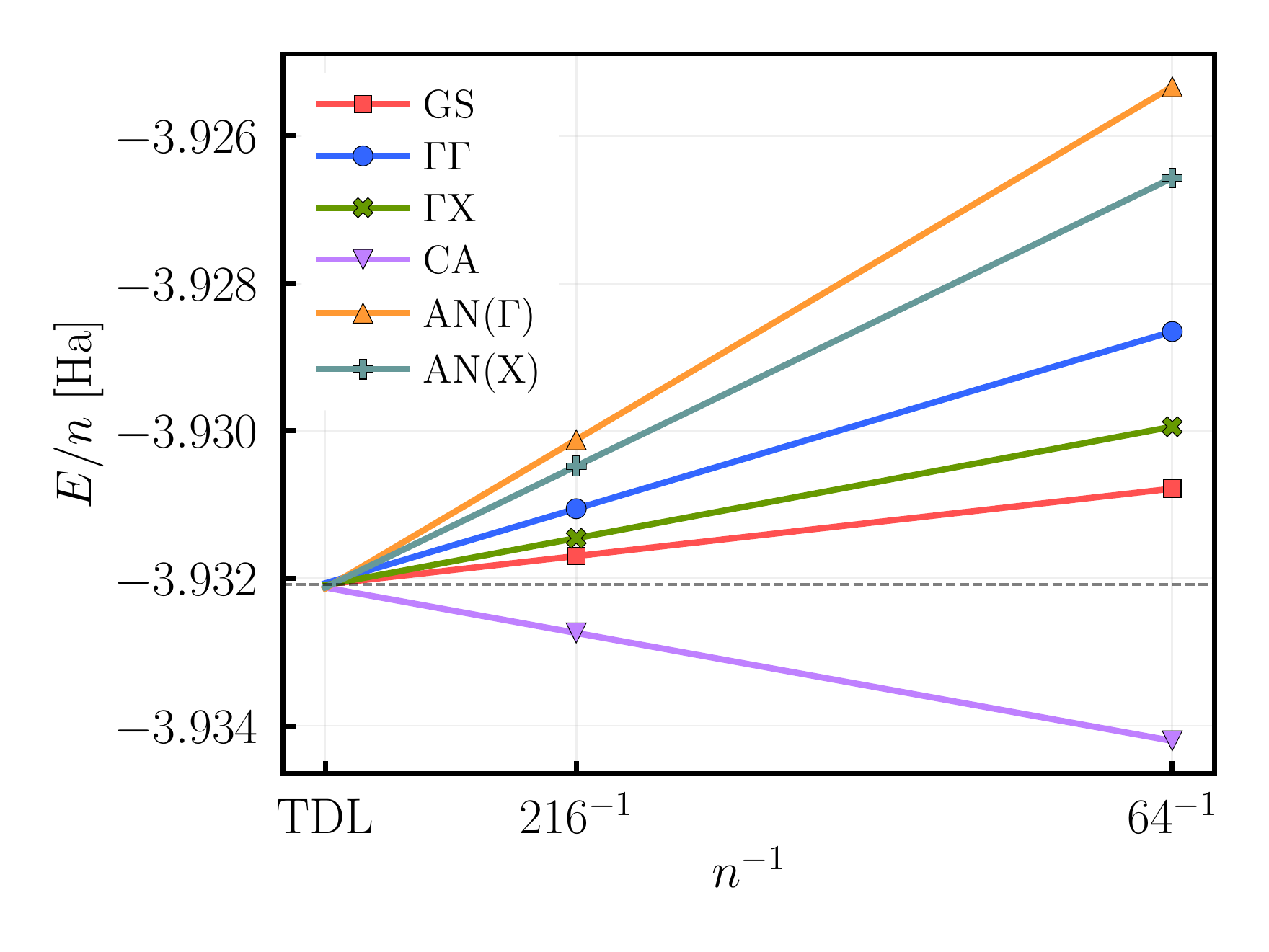}
\end{figure}

Datasets from Tables~\ref{tab:totals} and \ref{tab:qmc_charged} are used to probe for
differences of the thermodynamic limit (TDL)
for the energy per atom, denoted as $e_0$ in Eqns.~\ref{eqn:gs_scaling},~\ref{eqn:ex_scaling}. 
Linear extrapolations used for all six DMC/PBE sets are depicted in Fig.~\ref{fig:pbe_tdl}.
The consistency between excited and ground states
as well as for ionized cases is encouragingly tight
[the range of these energy values is $\approx2(1)$ meV].

Clearly, one can explore larger cells to diminish the discrepancies further, however, the overall small discrepancies provided a clear validation for subsequent analysis and enabled us to avoid further costly calculations.



\subsection{Cohesive energy}

One of the quantities that serves as an important indicator of the quality of the many-body method and the corresponding correlated wave functions is the cohesive energy. 
Table~\ref{tab:cohesion_HF} provides the cohesive energies for uncorrelated HF wave functions.
Note that our HF cohesion is $\sim 0.1$ eV better than previously reported independent calculations based on basis set expansions of many-body wave functions~\cite{mcclain_gaussian-based_2017, gruneis_second-order_2010} that so far were the most accurate known.

\begin{table}[!htbp]
\centering
\caption{
Hartree-Fock cohesive energies [eV] as obtained by the corresponding self-consistent codes and {by} VMC/HF (no Jastrow) in $\Gamma$-point occupation and extrapolation $n\to \infty$.
Complete basis set extrapolated ROHF atomic energy of $-3.6724778(1)$ Ha was used for the Si atom.
}
\label{tab:cohesion_HF}
\begin{tabular}{llr}
\hline
\rowcolor{lightgray!30}[2pt] Method & HF Cohesion [eV] & Ref. \\
\hline

HF/\textsc{crystal}      & 2.9912    & this work\footnote{Referenced to the exact atomic ROHF energy. \label{foot:HF_coh}} \\
VMC/HF/\textsc{crystal}  & 3.013(1)  & this work\textsuperscript{\ref{foot:HF_coh}} \\
HF/\textsc{QE}           & 3.1649    & this work\textsuperscript{\ref{foot:HF_coh}} \\
VMC/HF/\textsc{QE}       & 3.2018(6) & this work\textsuperscript{\ref{foot:HF_coh}} \\
& & \\

SCF(TZVP)                & 3.03      & McClain et al.\cite{mcclain_gaussian-based_2017} \\
SCF(PAW)\footnote{{PAW: Projector-augmented-wave method}}
                         & 2.97      & Grüneis et al.\cite{gruneis_second-order_2010} \\

\hline
\end{tabular}
\end{table}

Table~\ref{tab:cohesive_compare} provides the cohesive energy obtained in this work compared with previous calculations as well as with experiment.
We present the cohesive energies using the \mbox{DMC/PBE0} results since they correspond to the lowest obtained energies.
It is reassuring to see that nearly all DMC estimations of the cohesive energy agree with the experiment within the uncertainties regardless of the single-particle approach used for generating the orbitals.  Our results are represented by the following two estimations:

i) The first one is referenced to the exact Si atom [$-3.762073(57)$ Ha] as calculated by full-CI with complete basis set extrapolation.  

ii) The second estimator used the atomic fixed-node DMC energy [$-3.7601(1)$ Ha], resulting in partial error cancellation.  We consider this result to be the closest to the true value.

\begin{table}[!htbp]
\centering
\caption{
Cohesive energy [eV] obtained from DMC/PBE0 data compared with independent calculations and with the experimental value.
All values were corrected by zero-point energy \mbox{(0.06 eV)}~\cite{schimka_improved_2011} to correspond to the bottom of the interaction well ($D_e$).
}
\label{tab:cohesive_compare}
\begin{tabular}{llr}
\hline
\rowcolor{lightgray!30}[2pt] Method & Cohesion [eV] & Ref. \\
\hline

DMC/PBE0    & 4.629(2) & this work\footnote{Referenced to the exact Si atomic energy.} \\
DMC/PBE0    & 4.683(3) & this work\footnote{Referenced to the fixed-node DMC Si atomic energy.} \\
& & \\
DFT/LDA     & 5.1      & Dappe et al.\cite{dappe_local-orbital_2006} \\
MP2         & 5.05     & Grüneis et al.\cite{gruneis_second-order_2010} \\
MP2         & 4.96     & McClain et al.\cite{mcclain_gaussian-based_2017} \\
CCSD        & 4.15     & McClain et al.\cite{mcclain_gaussian-based_2017} \\
DMC/LDA     & 4.57(3)  & Li et al.\cite{li_cohesive_1991} \\
VMC/LDA     & 4.54(1)  & Leung et al.\cite{leung_calculations_1999} \\
DMC/LDA     & 4.69(1)  & Leung et al.\cite{leung_calculations_1999} \\
DMC         & 4.68(2)  & Alf\`e et al.\cite{alfe_diamond_2004} \\
& & \\
Experiment  & \textbf{4.68(8)}  & Farid et al.\cite{farid_cohesive_1991} \\

\hline
\end{tabular}
\end{table}

The comparison between these two estimators reveals an important insight into the systematic errors since the only difference is whether we account for the atomic fixed-node bias or not. 
Here the finite-size errors are significantly smaller, and since we use essentially a saturated basis set and very accurate ccECPs, the fixed-node error becomes the dominant remaining bias.
This suggests that a reasonable estimate of the total {\em systematic} error of our cohesive energy 
is approximately 0.05 eV. Consequently, we can write
\begin{equation}
E_{coh}= 4.683 \pm 0.05_{\rm (syst)} \pm 0.003_{\rm (stat)} \; {\rm eV}
\end{equation}
where the first deviation indicates the estimated systematic error while the second one corresponds to the statistical DMC error.
If we assume that $\approx$\mbox{4.68 eV} is the true value of the cohesive energy, that implies that the DMC solution of the many-body problem for this particular system is indeed very accurate 
with the correlation energy deficit of only
{
\mbox{$\eta = 1.3(2)\%$}, where $\eta$ is defined as:
\begin{equation}
    \eta = \frac{e_0^{\rm exact} - e_0^{\rm DMC}}{e_0^{\rm exact} - e_0^{\rm HF}} \cdot 100 \%.
\end{equation}
}
This also corroborates a very good agreement between independent DMC cohesive energy estimations. In addition, we can further infer that the Si crystal is very well described by the  single-reference trial wave function. This is a highly nontrivial result since otherwise we are not aware of any 
\textit{a priori} argument why this should be the case. 
Note that this is true regardless of the fact that the trial function corresponds to the direct product of nodes from the two spin-subspaces which is almost surely not correct.
The result also implies that the cancellation of errors with the atom is almost perfect since the FN error of the atom happens to be also almost identical \mbox{$\approx$ 0.054 eV}.
More support for these conjectures
comes from our recent 
 calculations of Si$_x$H$_y$ molecules where on average \mbox{$\sim 2\%$} fixed-node errors were observed for ground states and \mbox{$\sim 2.7\%$} for excited states which roughly corresponds to 0.05 eV. 

Of course, we do not expect such favorable error cancellation in general. In fact, it is more difficult to guarantee the same degree of accuracy for excited states since these are often more complicated due to possible multireference effects or other obstacles such as difficulties in obtaining the corresponding fully self-consistent orbital sets. 

\subsection{Quasiparticle and optical gaps}

For the sake of clarity, we define  
the optical gap as is customary, as the difference between ground and excited states:
\begin{equation}
    \label{eqn:promotion}
    {E_{g} = E^{EX}_{N} - E^{GS}_{N}}.
\end{equation}
The quasiparticle gap definition using cation, anion, and neutral systems is given by:
\begin{equation}
    \label{eqn:quasi-particle}
    {E_{G} = IP -EA = E^{AN}_{N+1} + E^{CA}_{N-1} - 2 \cdot E^{GS}_{N}}
\end{equation}
where $IP$ is the ionization potential, $EA$ is the electron affinity, and $N$ is the number of electrons.

Typically, 
both promotion and quasiparticle gaps
are calculated directly from the definition as differences of extensive total energies. Recently,
we suggested using slopes of intensive, per particle energies as less biased estimators. 
The {slopes of intensive energies as functions of $1/n$} enable us to enforce the common thermodynamic limit for both states 
and therefore partially diminish some biases in extensive energies~\cite{melton_many-body_2020}. 
We recast  Eqn.~\ref{eqn:gs_scaling} as follows:
\begin{equation}
    \label{eqn:gs_scaling_n}
    E_{N}^{GS}/n=e_0 + A^{GS}/n+g^{GS}(N)/n
\end{equation}
with analogous rearrangements for $E_N^{EX}$, $E^{CA}_{N-1}$, and $E^{AN}_{N+1}$.
In each of these energy expressions we neglect the last term that is approximated as
\begin{equation}
  g(N)/n \approx B/n^{\alpha+1}  
\end{equation}
where $B$ is a constant.
The gaps can be rewritten using slopes $S$ for each state as:
\begin{equation}
    \label{eqn:promotion_slope}
    \widetilde{E_{g}} = S^{EX}_{N} - S^{GS}_{N}
\end{equation}
where $S^{GS}_{N} = A^{GS}$ and $S^{EX}_{N} = (E_g+A^{EX})$.
Similarly, assuming that the offset constants approximately cancel out, one can express the fundamental gaps using slopes as follows
\begin{equation}
    \label{eqn:qp_slope}
    \widetilde{E_{G}} = S^{AN}_{N+1} + S^{CA}_{N-1} - 2 \cdot S^{GS}_{N}.
\end{equation}

Note that one can enforce the energy per atom $e_0$ in the thermodynamic limit to be identical for both ground and excited states.
We call this construction ``constrained-fit''.
It is also possible to keep values of $e_0$ as they are determined by an independent fit parameter for each state (see Fig.~\ref{fig:pbe_tdl}) and we refer to this as ``free-fit''.
More detailed discussions about obtaining gaps using slopes of intensive energies can be found in Ref.~\cite{melton_many-body_2020}.

{\em Spin contamination.} Nominally, an excited state which is constructed from a single-determinant with an electron promoted in one spin channel introduces spin contamination to the trial wave function, specifically, this state represents a \textit{mixture} of pure singlet and triplet states.
However, we found that the biases due to this issue are small or comparable to other systematic errors.
This is illustrated in Table~\ref{tab:spin_contamination} which shows the energies for the pure singlet, triplet, and mixed states.
Therefore, we used the mixed state with a single determinant for excited states throughout this work for simplicity.
Next, we present the gap estimations using PBE and PBE0 references as trial wave functions in VMC and DMC.

\begin{table}[!htbp]
\centering
\caption{
QMC/PBE0 total and kinetic energies [Ha] for $n=[8, 64]$ atoms in $\Gamma\Gamma$ excited state.
Triplet state is single-determinant WFN where an electron is transferred from one spin channel to the other.
Singlet state is a two-determinant WFN: $\Psi^{spatial}_{singlet} = \frac{1}{\sqrt{2}}\left(\alpha_{EX}\beta_{GS}+\alpha_{GS}\beta_{EX}\right)$.
Mixed state is a single-determinant WFN: $\Psi^{spatial}_{mixed} = \left(\alpha_{EX}\beta_{GS}\right)$.
The same Jastrow was used for all states above.
}
\label{tab:spin_contamination}
\begin{tabular}{cc|lllll}
\hline
\rowcolor{lightgray!30}[2pt] State & Method & Total & Kinetic \\
\hline
\rowcolor{lightgray!30}[2pt] \multicolumn{4}{c}{$n=8$} \\
\hline
Singlet & VMC & -31.0462(3) & 14.0271(10) \\
Mixed   & VMC & -31.0458(2) & 14.0289(10) \\
Triplet & VMC & -31.0498(2) & 14.0272(10) \\
\hline
Singlet & DMC & -31.1343(4) & 14.0193(16) \\
Mixed   & DMC & -31.1360(4) & 14.0158(15) \\
Triplet & DMC & -31.1396(4) & 14.0149(15) \\
\hline
\rowcolor{lightgray!30}[2pt] \multicolumn{4}{c}{$n=64$} \\
\hline
Singlet & VMC & -251.0356(3)  & 107.3878(17) \\
Mixed   & VMC & -251.0358(3)  & 107.3880(12) \\
Triplet & VMC & -251.0363(3)  & 107.3917(14) \\
\hline
Singlet & DMC & -251.4331(17) & 107.352(12)  \\
Mixed   & DMC & -251.4371(10) & 107.341(12)  \\
Triplet & DMC & -251.4381(11) & 107.3672(86) \\
\hline
\end{tabular}
\end{table}


{\em Gaps with PBE orbitals. }
The band gaps from extensive energies with PBE orbitals are given 
in Table~\ref{tab:extensive_pbe_gaps}.
On the other hand, Table~\ref{tab:intensive_pbe_gaps} presents the gaps estimated from slopes with 
the types of constructions introduced above. We can see that the agreement 
is very good in general, with better consistency between VMC and DMC estimations using the slopes
and the constraints. 
In particular, when $e_0$ values show minor differences, constraining the values of $e_0$ reduces the minor biases
as also observed before ~\cite{melton_many-body_2020}.
The explicit data for each $n$ is provided in the Supplemental Material~\cite{supplemental}.
Using the slope estimators, we can see that there is a good agreement between promotion and quasiparticle gaps within the error bars (Table~\ref{tab:intensive_pbe_gaps}), 
so that we can write:
\begin{equation}
    \label{eqn:fund_optical}
    E_{G} \approx E_{g}.
\end{equation}
This is expected for the Si crystal since
energy wise an exciton in larger supercells should be significantly below 0.1 eV
\cite{green_improved_2013}.
Indeed, eight atom supercells show some deviations between promotion vs quasiparticle gaps, but this difference disappears in larger supercells. 

\begin{table}[!htbp]
\centering
\caption{
QMC/PBE gaps [eV] using extensive total energies as in Eqns.~\ref{eqn:promotion},~\ref{eqn:quasi-particle} for $n=[8,64,216]$ atoms.
}
\label{tab:extensive_pbe_gaps}
\begin{tabular}{ccl|lll}
\hline
\rowcolor{lightgray!30}[2pt] Gap & Method & WFN & $n=8$ & $n=64$ & $n=216$ \\
\hline
\rowcolor{lightgray!30}[2pt] \multicolumn{6}{c}{Promotion gaps $E_g$} \\
\hline
$\Gamma\Gamma$ & VMC & S  &  3.61(2) &   5.1(1) &   5.7(1) \\
$\Gamma\Gamma$ & VMC & SJ &  3.12(2) &  3.68(4) &  3.71(5) \\
$\Gamma\Gamma$ & DMC & SJ &  3.03(4) &  3.71(9) &   3.8(1) \\
\hline
$\Gamma$X      & VMC & S  &  1.35(2) &  2.55(9) &  3.39(8) \\
$\Gamma$X      & VMC & SJ &  0.94(2) &  1.36(4) &  1.49(4) \\
$\Gamma$X      & DMC & SJ &  0.84(4) &  1.46(8) &   1.4(2) \\
\hline
\rowcolor{lightgray!30}[2pt] \multicolumn{6}{c}{Quasiparticle gaps $E_G$} \\
\hline
$\Gamma\Gamma$ & VMC & S  &  3.54(4) &   4.9(2) &   5.6(1) \\
$\Gamma\Gamma$ & VMC & SJ &  3.04(3) &  3.67(7) &  3.75(7) \\
$\Gamma\Gamma$ & DMC & SJ &  2.96(6) &   3.5(1) &   3.1(2) \\
\hline
$\Gamma$X      & VMC & S  &  1.19(4) &   2.4(2) &   3.2(1) \\
$\Gamma$X      & VMC & SJ &  0.75(3) &  1.37(7) &  1.47(7) \\
$\Gamma$X      & DMC & SJ &  0.61(6) &   1.4(1) &   1.0(2) \\
\hline
\end{tabular}
\end{table}

\begin{table}[!htbp]
\centering
\caption{
QMC/PBE gaps [eV] using intensive energies and slopes as in Eqns.~\ref{eqn:promotion_slope},~\ref{eqn:qp_slope} for free and constrained fits with 64-216 atom extrapolation. 
}
\label{tab:intensive_pbe_gaps}
\begin{tabular}{ccl|lll}
\hline
\rowcolor{lightgray!30}[2pt] Gap & Method & WFN & free-fit & const-fit \\
\hline
\rowcolor{lightgray!30}[2pt] \multicolumn{5}{c}{Promotion gaps $E_g$} \\
\hline
$\Gamma\Gamma$ & VMC & S  &   4.9(1) &   5.4(3) \\
$\Gamma\Gamma$ & VMC & SJ &  3.66(7) &  3.69(2) \\
$\Gamma\Gamma$ & DMC & SJ &   3.7(1) &  3.74(3) \\
\hline
$\Gamma$X      & VMC & S  &   2.2(1) &   3.1(4) \\
$\Gamma$X      & VMC & SJ &  1.31(6) &  1.42(6) \\
$\Gamma$X      & DMC & SJ &   1.5(1) &  1.46(1) \\
\hline
\rowcolor{lightgray!30}[2pt] \multicolumn{5}{c}{Quasiparticle gaps $E_G$} \\
\hline
$\Gamma\Gamma$ & VMC & S  &   4.7(3) &   5.4(2) \\
$\Gamma\Gamma$ & VMC & SJ &   3.6(1) &  3.69(3) \\
$\Gamma\Gamma$ & DMC & SJ &   3.7(2) &   3.4(1) \\
\hline
$\Gamma$X      & VMC & S  &   2.1(3) &   2.9(3) \\
$\Gamma$X      & VMC & SJ &   1.3(1) &  1.41(4) \\
$\Gamma$X      & DMC & SJ &   1.5(2) &   1.3(1) \\
\hline
\end{tabular}
\end{table}

\begin{table}[!htbp]
\centering
\caption{
QMC/PBE0 gaps [eV] using extensive total energies as in Eqn.~\ref{eqn:promotion} for $n=[8,64,216]$ atoms.
}
\label{tab:extensive_pbe0_gaps}
\begin{tabular}{ccl|lll}
\hline
\rowcolor{lightgray!30}[2pt] Gap & Method & WFN & $n=8$ & $n=64$ & $n=216$ \\
\hline
\rowcolor{lightgray!30}[2pt] \multicolumn{6}{c}{Promotion gaps $E_g$} \\
\hline
$\Gamma\Gamma$ & VMC & S  &  3.60(2) &  5.35(6) &  6.17(5) \\
$\Gamma\Gamma$ & VMC & SJ &  3.13(2) &  3.64(1) &  3.78(2) \\
$\Gamma\Gamma$ & DMC & SJ &  2.94(3) &  3.74(4) &   4.0(1) \\
\hline
$\Gamma$X      & VMC & S  &  1.20(2) &  2.84(5) &  3.61(6) \\
$\Gamma$X      & VMC & SJ &  0.80(3) &  1.38(2) &  1.48(2) \\
$\Gamma$X      & DMC & SJ &  0.61(3) &  1.49(5) &   1.5(1) \\
\hline
\end{tabular}
\end{table}

\begin{table}[!htbp]
\centering
\caption{
QMC/PBE0 gaps [eV] using intensive energies and slopes as in Eqn.~\ref{eqn:promotion_slope} for free and constrained fits with 64-216 atom extrapolation. 
}
\label{tab:intensive_pbe0_gaps}
\begin{tabular}{ccl|lll}
\hline
\rowcolor{lightgray!30}[2pt] Gap & Method & WFN & free-fit & const-fit \\
\hline
\rowcolor{lightgray!30}[2pt] \multicolumn{5}{c}{Promotion gaps $E_g$} \\
\hline
$\Gamma\Gamma$ & VMC & S  &   5.00(8) &   5.7(4) \\
$\Gamma\Gamma$ & VMC & SJ &   3.59(2) &  3.67(5) \\
$\Gamma\Gamma$ & DMC & SJ &   3.63(8) &  3.77(8) \\
\hline
$\Gamma$X      & VMC & S  &   2.52(8) &   3.0(3) \\
$\Gamma$X      & VMC & SJ &   1.34(3) &  1.44(5) \\
$\Gamma$X      & DMC & SJ &   1.46(8) &  1.50(2) \\
\hline
\end{tabular}
\end{table}

{\em Gaps with PBE0 orbitals.   }
For the case of PBE0 orbitals, we calculated only promotion
band gaps since we expect general agreement as observed above for PBE orbitals. Note that there are also (perhaps minor) technical advantages in favor   of promotion gaps. One of these is the error bars are 
smaller in general (due to difference of two total energies instead of multiple ones for quasiparticle gaps). In addition, the charged supercells
show a tendency to enhance the systematic biases as discussed further in the next section. 
This is also visible in Fig.~\ref{fig:pbe_tdl} that clearly shows that the slopes are the largest for charged cells.  

In Table~\ref{tab:extensive_pbe0_gaps} we list the \mbox{QMC/PBE0} gaps using extensive energies while in Table~\ref{tab:intensive_pbe0_gaps} the same gaps are estimated from  slopes. This latter set we consider as our most consistent and accurate results.
As a summary, Table~\ref{tab:gaps_compare} presents these results  compared to other independent calculations and experiments.
Our results show a notable improvement over previous calculations; however, there appears to be a minor $\sim0.2$ eV overestimation of gaps. 
We identify a couple of most plausible possibilities for this overestimation:
\begin{enumerate}
    \item The FN error cancellation is not perfect for ground and excited states - this has been observed also in small Si clusters~\cite{wang_binding_2020}.
    Another related point is that the single-particle orbitals are optimized for the ground state and therefore they are not fully relaxed for excited states.
    We note that direct optimization of the orbitals and wave functions for both the ground and excited states has been carried out previously for band gap calculations~\cite{neuscamman_2019}. However, this approach is currently limited to small supercells while we are interested in obtaining band gaps in the thermodynamic limit. Therefore, we are limited to trial wave functions built from the mean-field orbitals. 
    \item The terms $A^{GS}$, $A^{EX}$ in Eqns.~\ref{eqn:gs_scaling},~\ref{eqn:ex_scaling} do not necessarily cancel out. 
    Namely, the difference $\Delta_A = A^{EX} - A^{GS}$ will persist as $\mathcal{O}(1)$ constant even for large $n$ values.
    Using intensive energies  does not eliminate this particular problem since the slope difference is \mbox{($E_g + \Delta_A$)} so that the bias from offsets ``sticks'' to the gap value.
\end{enumerate}

{
Additionally, Table~\ref{tab:gaps_compare} shows that \mbox{VMC/PBE0 (SJ)} provides the same quality gaps as DMC/PBE0 (SJ).
We conclude that for this system with cubic shape of supercells, \mbox{VMC/DFT (SJ)} could be used for future gap studies with significant computational savings.
}

\begin{table}[!htbp]
\centering
\caption{
Gaps [eV] obtained in this work using extrapolated DMC/PBE0 data compared with previous independent
calculations and with experimental values. 
Experimental gaps were increased by a zero-point band gap renormalization value of \mbox{64 meV} 
\cite{cardona_isotope_2005} (see also Ref.~\cite{miglio_predominance_2020}).
}
\label{tab:gaps_compare}
\begin{tabular}{lllr}
\hline
\rowcolor{lightgray!30}[2pt] Method & WFN & Gap [eV] & Ref. \\
\hline
\rowcolor{lightgray!30}[2pt] \multicolumn{4}{c}{$\Gamma\Gamma$ } \\
\hline
DFT/PBE0 & S  & 3.96      & this work \\
VMC/PBE0 & SJ & 3.67(5)   & this work \\
DMC/PBE0 & SJ & 3.77(8)   & this work \\
& & & \\
$GW$     &    & 3.32      & Rieger et al.\cite{rieger_gw_1999} \\
DMC/PBE  & SJ & 4.14(3)   & Hunt et al.\cite{hunt_quantum_2018} \\
& & & \\
Experiment & & \textbf{3.44}      & Jellison et al.~\cite{jellison_optical_1983} \\
\hline
\rowcolor{lightgray!30}[2pt] \multicolumn{4}{c}{$\Gamma$X } \\
\hline
DFT/PBE0 & S  & 1.84      & this work \\
VMC/PBE0 & SJ & 1.44(5)   & this work \\
DMC/PBE0 & SJ & 1.50(2)   & this work \\
& & & \\
$GW$     &    & 1.42      & Rieger et al.\cite{rieger_gw_1999} \\
DMC/PBE  & SJ & 1.9(1)    & Hunt et al.\cite{hunt_quantum_2018} \\ 
DMC      & BF\footnote{{BF: backflow wave function}}
              & 1.7(1)    & Yang et al.\cite{yang_electronic_2020} \\
& & & \\
Experiment & & \textbf{1.31}      & Ortega et al.\cite{ortega_inverse-photoemission_1993} \\
\hline
\end{tabular}
\end{table}

\subsection{Estimation of IP and EA for Si crystal}

Using the energies from Table~\ref{tab:totals} and~\ref{tab:qmc_charged}, we can evaluate $IP$ and $EA$ from the extensive energies as given by
\begin{equation}
    IP = E^{CA}_{N-1} - E^{GS}_{N}
\end{equation}
\begin{equation}
    EA = E^{GS}_{N} - E^{AN}_{N+1}.
\end{equation}
Clearly, we run into a problem since the raw values do not give meaningful results, see Table~\ref{tab:totals}.
They are negative while for a stable system they must be positive.
The reason for this naively incorrect result is the well-known nonuniqueness of total energy for charged periodic systems recognized long time ago, see, for example, Refs.~\cite{persson_n-type_2005,lany_assessment_2008} and references therein. 
Note that the charged supercell energy can be shifted by some effective chemical potential, which results from a particular balance between kinetic and potential energy components given by the adopted potential energy and periodicity model. Related issues such as  offsets of eigenvalues
as well as nonuniqueness of total energy of charged periodic systems are present also in DFT calculations~\cite{lany_assessment_2008, persson_n-type_2005}. 

In order to sort this out, one needs to define the reference (zero) level of the potential appropriately. The proper reference is the vacuum level at some point infinitely far from the considered system since this 
corresponds to relevant experiments such
as direct or inverse photoemissions. 
We therefore define the Fermi level $E_{FL}$ as customary for intrinsic semiconductors to be in the center of the band gap
\begin{equation}
    \label{eqn:fermi_level}
    E_{FL} = -(IP + EA)/2 = -(E^{CA}_{N-1} - E^{AN}_{N+1})/2
\end{equation}
where $IP$ and $EA$ values are top/bottom energies of the corresponding valence/conduction bands referenced to vacuum.
Note that this would be correct assuming our supercell total energies would be also referenced accordingly.
For isolated systems such as atoms or molecules in vacuum this level is naturally defined by the zero of the Coulomb potential at infinity. 
However, our model of potential energy and corresponding Ewald sums together with an imperfect balance with the other energy components produce an offset.
Unless compensated, this offset survives to the thermodynamic limit. Unfortunately, this issue is further complicated by local effects from core states (or by effective core potentials that mimic the core states), finite size supercell, $k-$point occupation, and also by the
correlation treatment level. 
In order to take this nominally unknown shift into account, we write the supercell Fermi level offset by a constant $\Delta_s$
\begin{equation}
    \label{eqn:fermi_shift}
    \tilde E_{FL} = E_{FL}+ \Delta_s 
\end{equation}
Now we can express the cation ($q=1$, hole) and the anion ($q=-1$, electron) supercell  total 
energies as follows:
\begin{equation}
    \label{eqn:ca_fermi}
    E_{N-1}=E_N+E_G/2 -q\tilde E_{FL}\quad (q=1)\;\;\;
\end{equation}
\begin{equation}
    \label{eqn:an_fermi}
    E_{N+1}=E_N+E_G/2 -q\tilde E_{FL}\quad (q=-1)
\end{equation}

\begin{table*}[!htbp]
\centering
\caption{
QMC/PBE estimation of Fermi level $\tilde E_{FL}$ [eV] using raw total energies   $\tilde E_{FL} = -(E^{CA}_{N-1} - E^{AN}_{N+1})/2$ from both extensive  and extrapolation formulations. Note that $\tilde E_{FL}$ includes an artificial offset as discussed in text, 
Eqn. \ref{eqn:fermi_shift}.
}
\label{tab:an_ca_gaps}
\begin{tabular}{ccl|lllll}
\hline
\rowcolor{lightgray!30}[2pt] Qty. & Method & WFN & 8 & 64 & 216 & $\infty_{\rm free-fit}$ & $\infty_{\rm const-fit}$ \\
\hline
(AN($\Gamma)-$CA)/2 & VMC & S  & 8.875(7) &   9.01(5) &  9.09(3) &  8.97(7) &   9.1(1) \\
(AN($\Gamma)-$CA)/2 & VMC & SJ & 7.826(6) &  7.758(9) &  7.77(2) &  7.75(1) &  7.76(1) \\
(AN($\Gamma)-$CA)/2 & DMC & SJ &  7.78(2) &   7.73(4) &  7.70(6) &  7.74(7) &  7.72(7) \\
\hline
(AN(X)$-$CA     )/2 & VMC & S  & 7.701(7) &   7.74(5) &  7.90(4) &  7.67(7) &   7.8(1) \\
(AN(X)$-$CA     )/2 & VMC & SJ & 6.684(6) &   6.61(1) &  6.63(2) &  6.60(2) &  6.63(2) \\
(AN(X)$-$CA     )/2 & DMC & SJ &  6.60(2) &   6.65(4) &  6.65(7) &  6.65(7) &  6.66(6) \\
\hline
\end{tabular}
\end{table*}

One could argue that the shift is not necessarily constant and that it could vary with the supercell size. 
However, this is not the case and in fact, the shifted Fermi level $\tilde E_{FL}$ is remarkably constant as can be seen from Table~\ref{tab:an_ca_gaps} (using energies from Table~\ref{tab:totals} and~\ref{tab:qmc_charged}). 
This behavior holds for the expectation energy of the bare Slater determinant, as well as for the VMC and DMC methods, even for a very small supercell size with eight atoms. Interestingly, it exhibits  smaller variation than, for example,
the band gaps from differences of total energies listed in the tables above.
The early onset of the Fermi level invariance on size suggests that it should be possible to estimate it from related smaller systems. In particular,
Si clusters with  atoms in similar bonding patterns show Fermi levels that are comparable to the bulk~\cite{haberlen_clusters_1997, kikuchi_first-principles_2007}. Perhaps even 
more surprisingly, free-standing 
Si clusters of very small sizes such as Si$_6$ - Si$_{11}$ show mildly varying Fermi levels that are very close to the {\em atomic Fermi level} given as an average of EA and IP~\cite{tam_heats_2013}. This is true despite the fact that IP and EA values themselves change by several
eVs from their atomic values. A similar trend is observed 
for larger, hydrogen saturated clusters 
\cite{melnikov_electron_2004, zhou_electronic_2003}. Although the convergence is not monotonous due to the shell effects in cluster geometries and the varying number of terminating atoms, the tendency towards 
the bulk values of band gap, Fermi level, and work function
are unmistakable. There are basically two key reasons for observing these trends:

a) First,  
both the clusters and the Si bulk
are monoatomic systems  with nonpolar bonds
and closed-shell ground states. Absence of charge transfers as well as presence of gaps therefore incur significant constraints on restructuring of the energy levels. 

b) Second, note that electron affinities and ionization potentials involve predominantly, and for the considered states almost exclusively, only the $p$ levels. In the solid the valence band maximum $\Gamma_{25'}$  as well as the X-band conduction X$_{1c}$ states are essentially $p$ bands. Note that the same applies to the atom where affinity and ionization comes from 
changes in occupations of the $p$ subshell. This implies that a $p$-band model with symmetric electron-hole levels should be an appropriate picture of the ionized states, both in the atom and in the solid.
Indeed, similar electron-hole
symmetry was  found in related systems by many-body perturbation methods such as GW~\cite{tamblyn_electronic_2011, jiang_ionization_2013}.

Consequently,
while the relevant levels move very significantly from the constituent atom to the insulating bulk, we assume that these shifts are symmetric with regard to the Fermi level.
Therefore, for our system, we estimate the position of the Fermi level referenced to the vacuum by its atomic value:
\begin{equation}
    \label{eqn:atom_solid_Ef}
    E_{FL}^{\rm \; Si \; solid} \approx E_{FL}^{\rm \; Si\; atom}
\end{equation}
which we take from the nearly exact atomic ccECP calculations
\cite{bennett_new_2018} (see Supplemental Material~\cite{supplemental}):
\begin{equation}
    E_{FL}^{\rm \; Si \; atom,\; exact} = -0.1759 \; \textrm{Ha}
\end{equation}
Considering Eqns.~\ref{eqn:quasi-particle},~\ref{eqn:fund_optical},~\ref{eqn:fermi_level},~\ref{eqn:atom_solid_Ef} we get familiar  expressions for intrinsic semiconductor  $IP$ and $EA$ given by
\begin{equation}
    IP = E_g/2 - E_{FL}, \quad
    EA = - E_g/2 - E_{FL}
\end{equation}
where $E_g$ represents the actual conduction band minimum (CBM) $-$ valence band maximum (VBM) gap.
Furthermore, for $E_g$ we use our DMC/PBE0 estimates of $E_g$($\Gamma$X) corrected by a small  value of $\approx$0.08 eV which corresponds to both experimentally and theoretically known difference between CBM$-$VBM gap and $\Gamma$X gap~\cite{jellison_optical_1983, ortega_inverse-photoemission_1993}.
Based on these considerations,
Table~\ref{tab:ip_ea_compare} provides the estimations for $IP$ and $EA$ using the $E_{FL}$ and $E_g$ values. 
Remarkably, even this tentative assessment leads to very reasonable $IP$ and $EA$ values that compare favorably both with other calculations and with experiments.
We note that while our estimations are for an ideal crystal, the experiments involve possible surface effects that we do not consider here at all.
It is also clear that for other systems the Fermi level might be more complicated to find, for example,
by using calculations of a surface or a slab to properly align the corresponding energy levels or other approaches
(see, for example, Refs.~\cite{lany_assessment_2008, persson_n-type_2005}
and papers cited therein).

In our definitions above, the total energies of charged systems are shifted by a {\em constant} addition~\cite{lany_assessment_2008} to the Fermi energy/chemical potential, which is the same regardless of whether an electron was added or subtracted.
We can estimate $\Delta_s$ using the DMC/PBE data:
\begin{equation}
    \Delta_s = \tilde E_{FL} - E_{FL} \approx 0.2446 + 0.1759 = 0.4205 \; {\rm Ha} 
\end{equation}
where $\tilde E_{FL}$ corresponds to $k=$ X occupied anion state and $\infty_{\rm const-fit}$ value.
This offset is substantial and it overshadows the true value of the Fermi level with consequences that we described before. 
Note that in the calculations of fundamental gaps, the Fermi level (true or shifted) cancels out so
that the gap calculations are not affected. 

The precise value of the offset does not have a single source
and therefore it is not straightforward to identify its genuine origin as we already alluded to above. Its large value, at least in our definition, 
suggests that the localized atomic contributions are dominant. Its size invariance points toward an energy density contribution that can be also recast as corresponding effective chemical potential(s)
\cite{yang_electronic_2020, azadi_efficient_2019, krogel_quantum_2013}. 
This is further tied to
 the oversimplified charge compensation by the constant background and associated contributions generated by the artificial periodicity. 
The constant background is perfectly appropriate for the homogeneous electron gas since its 
density is constant as well. On the other hand, for inhomogeneous systems the electrostatic model should be more elaborate in order to not only cancel out the divergences but also to counteract any related finite offset. Many-body effects such as exchange and correlation add another facet to this. Note that an appropriate model might involve also state and correlation treatment dependencies as suggested by the variation of the shift between
methods and chosen states, see
Table \ref{tab:an_ca_gaps}.  This clearly
calls for further elaboration in the future.

\begin{table}[!htbp]
\centering
\caption{
Estimations of IP and EA [eV] obtained in this work using $E_g$ gaps (see text) compared with experimental values and independent calculations.
}
\label{tab:ip_ea_compare}
\begin{tabular}{llr}
\hline
\rowcolor{lightgray!30}[2pt] Method & Qty. [eV] & Ref. \\
\hline
\rowcolor{lightgray!30}[2pt] \multicolumn{3}{c}{IP} \\
\hline
VMC/PBE0 & 5.46(2) & this work \\
DMC/PBE0 & 5.49(1) & this work \\
& & \\
$GW$(VBM)/PBE & 5.45 & Jiang et al.\cite{jiang_ionization_2013} \\ 
& & \\
Experiment &  5.10    & Gobeli et al.\cite{gobeli_photoelectric_1965} \\
Experiment &  5.35(2) & Sebenne et al.\cite{sebenne_surface_1975} \\
\hline
\rowcolor{lightgray!30}[2pt] \multicolumn{3}{c}{EA} \\
\hline
VMC/PBE0 & 4.11(2) & this work \\
DMC/PBE0 & 4.08(1) & this work \\
& & \\
$GW$(CBM)/PBE & 4.34 & Jiang et al.\cite{jiang_ionization_2013} \\ 
& & \\
Experiment  & 4.01 & Gobeli et al.\cite{gobeli_photoelectric_1965} \\
\hline
\end{tabular}
\end{table}

\section{Conclusions}
\label{section:conclusions}
We present real space QMC calculations of Si crystal which study cohesion, optical and fundamental gaps, and provide estimations of the ionization potential and the electron affinity.   
We emphasize considerations of systematic errors and the necessity of their analysis for reliable predictions.
Our data shows the importance  of basis set accuracy as well as probing of effective single-particle theories to generate the most optimal orbitals. 
Further errors involve the fixed-node biases and extrapolations from finite sizes that require sufficiently large supercells in order to reach a regime that is adequately close to the thermodynamic limit. We demonstrate that we reached this regime by consistently extrapolating to the same cohesive energy per atom for all calculated states, including ground and excited states with one-particle promotions as well as cation and anion states.  
We find that the optical and fundamental gaps agree with very good accuracy as observed also in previous calculations 
\cite{wines_first-principles_2020, melton_many-body_2020, dubecky_fundamental_2020}.
Overall,
only small gap discrepancies of the order of 0.2 eV were revealed when compared with experiments.
These residual errors are attributed to the remaining imperfections both from excited state trial functions that are marginally worse than the ground state ones as well as from probable residual finite size effects that need more refined estimations.

The calculations enabled us to estimate the cohesive energy with {\em both systematic and random errors}
under $\approx0.05$ eV, making this prediction fully {\em ab initio} with control over the remaining minor biases. In turn, the results suggest 
remarkably high accuracy of the ground state trial functions that provide {$98.7(2)\%$} of the correlation energy in fixed-node DMC method. This value is estimated by indirect comparisons with relevant smaller Si systems as well as by using the experiment as 
an additional indicator.

In general, the ionized states provide information about the electron affinity and ionization potential of the Si crystal. However, referencing the band edges to vacuum assumes that one can estimate the Fermi level
in bulk with desired accuracy. We used the atomic Fermi level for the intrinsic ideal crystal by providing arguments why this is appropriate (single homopolar bonds, monoatomic system, and the fact that relevant atomic and crystal states involve $p$ orbitals/$p$ bands only so that one-band model with electron-hole symmetry applies). The obtained electron affinity and ionization potential are in very good agreement with the experimental values
despite the fact that we did not consider any surface effects,
possibly suggesting that they might not play a major role for the Si solid.

We have analyzed the results for the charged states that are compensated 
by the homogeneous background within conventional Ewald summation techniques.
We have estimated the artificial offsets of the charged states and show how they obscure the true Fermi level by having large values with opposite signs. The offsets are essentially perfectly constant and result from  local atomic effects combined with oversimplified
charge compensation and periodicity model.

We would like to conclude with two key messages. Overall, the results suggest that the QMC methods are making systematic progress in addressing
much more subtle aspects of electron-electron correlation effects than, say, a decade ago.
At the same time, further analysis is needed to find more robust and more straightforward approaches to deal with some of the remaining technical biases 
that contaminate accurate QMC results.
 The significant progress that has been achieved is due to new insights into the nature of many-body effects, development of more sophisticated methods
as well as availability of new computational tools. 

\bigskip
\bigskip
\bigskip

{\bf Acknowledgements.} 
The authors would like to thank Jaron T. Krogel for the kind help with \textsc{nexus}.
We also thank Paul R. C. Kent and Raymond C. Clay III for reading the paper and helpful suggestions and comments. 
The presented work used has been funded by the U.S. Department of Energy, Office of Science, Basic Energy Sciences, Materials Sciences and Engineering Division,  as  part  of  the  Computational  Materials  Sciences Program and Center for Predictive Simulation of Functional Materials. 
An award of computer time was provided by the Innovative and Novel Computational Impact on Theory and Experiment (INCITE) program. This research used resources of the Argonne Leadership Computing Facility, which is a DOE Office of Science User Facility supported under contract DE-AC02-06CH11357.
This research also used resources of the Oak Ridge Leadership Computing Facility, which is a DOE Office of Science User Facility supported under Contract DE-AC05-00OR22725.
This research used resources of the National Energy Research Scientific Computing Center (NERSC), a U.S. Department of Energy Office of Science User Facility located at Lawrence Berkeley National Laboratory, operated under Contract No. DE-AC02-05CH11231.

Sandia National Laboratories is a multimission laboratory managed and operated by National Technology \& Engineering Solutions of Sandia, LLC, a wholly owned subsidiary of Honeywell International Inc., for the U.S. Department of Energy’s National Nuclear Security Administration under contract DE-NA0003525.

This paper describes objective technical results and analysis. Any subjective views or opinions that might be expressed in the paper do not necessarily represent the views of the U.S. Department of Energy or the United States Government.

\bigskip
{\bf Data Availability.} 
The input/output files and supporting data generated in this work are published in Materials Data Facility~\cite{blaiszik_materials_2016, blaiszik_data_2019} and can be found in Ref.~\cite{mdf_data}.
More information such as employed geometry can be found in Supplemental Material~\cite{supplemental}.

\newpage
\bibliography{main}

\end{document}